\documentclass[12pt]{article}
\usepackage{amsmath}
\usepackage{graphicx}
\usepackage{subfigure}
\usepackage{hyperref}

\begin{document}

\author{Ernst Trojan and George V. Vlasov \and \textit{Moscow Institute of Physics
and Technology} \and \textit{PO Box 3, Moscow, 125080, Russia}}
\title{Shock waves in superconducting cosmic strings: instability to extrinsic
perturbations}
\maketitle

\begin{abstract}
Superconducting cosmic string may admit shock-like discontinuities of the current when the latter is spacelike ("magnetic" regime), while no shock at timelike current ("electric" regime) was discovered in numerical simulations. 
We find that the necessary and enough conditions for existence of stable shocks and show that the shock can be unstable in the presence of infinitesimal extrinsic perturbations of the string worldsheet. 
The shocks in the "magnetic" regime are not vulnerable to this instability but the shocks in the "electric" regime do not survive.
\end{abstract}

%{\large

\sloppy

\section{Introduction}

Cosmic strings are 2-dimensional topological defects that are believed to be
formed at a phase transition in the early universe \cite{Kibble80,Vilenkin85}%
. When the core of the string is small with respect to its radius, the
behavior is determined by the Goto-Nambu action 
\begin{equation}
S=\int \Lambda \sqrt{\left| \det h_{ab}\right| }d\sigma ^1d\sigma ^2
\label{action}
\end{equation}
with the surface metric $h_{ab}$ and surface Lagrangian $\Lambda $. The
cosmic strings can be endowed with internal structure \cite{W85}, and their
Lagrangian is dependent on the magnitude of the current $\chi $, giving $%
\Lambda =-m^2$ in the limit $\chi \rightarrow 0$. Such superconducting
strings  admit a large variety of stable loop solutions (or, vortons) and
their intercommutation may frequently take place. The problem of vorton
stability and evolution of superconducting string networks attract the
interest of researchers because such string loops can accumulate significant
part of the universe mass and the expected dominant effect of their
reconnection is particle radiation (when some particles are expelled away).

The equations of motion admit solutions in the form of infinitesimal
perturbations of two types \cite{Carter89a}: extrinsic perturbations of the
world sheet which concern the string geometry,\ and sound type longitudinal
perturbations within the world sheet whcih concern the current $\chi $.
Nonlinear effects become dominant rapidly in longitudinal perturbations of
finite amplitude $\Delta \chi =\chi _{+}-\chi _{-}\neq 0$ and they may
either disappear or form stable discontinuities (magnitude $\chi $ becomes
discontinuous) similar to shock waves propagating along the string. Such
strings loops may tend to fold on themselves and make contact points of
self-intersection so that it will be energetically favored to some of the
trapped particles to move out of the string, thus, generating emission,
associated with possible visible consequences \cite
{Kibble80,Vilenkin85,BHV2001,FV2006}.

First, the shocks they were predicted \cite{V99} at spacelike currents $\chi
>0$ (''magnetic'' regime) and no shock was expected at timelike currents
(''electric'' regime) that was was referred to the evolutionary condition 
\cite{V99}. However, the latter imposes no restriction except the growth of
the current $\chi _{+}-\chi _{-}>0$ that can be satisfied in the
''electric'' regime as well \cite{TV2011a}. Nevertheless, according to
numerical simulations \cite{MP00,CCMP02}, the dynamical evolution of a
bosonic current carrier can develop shocks in the ''magnetic'' regime and no
discontinuous solution was discovered in the ''electric'' regime. As a
matter of fact, the ''electric'' shock waves are forbidden, and the problem
of shock stability at timelike currents remains unresolved.

In the present paper we analyze the phenomeneon which is responsible for the
absence of shocks in the ''electric'' regime. It is the shock instability
caused by perturbations of the string worldsheet.

\section{Intrinsic and extrinsic perturbations}

The dynamics of current-carying cosmics strings is determined by
''extrinsic'' equations of motion \cite{Carter89a} 
\begin{equation}
\perp _\sigma ^\mu \left( Uu^\nu \nabla _\nu \,u^\sigma -Tv^\nu \nabla _\nu
\,v^\sigma \right) =0\qquad \perp _\sigma ^\mu \left( u^\nu \nabla _\nu
v^\sigma -v^\nu \nabla _\nu \,u^\sigma \right) =0  \label{ex1}
\end{equation}
and ''intrinsic'' equations of motion 
\begin{equation}
\eta _\mu ^\nu \nabla _\nu \left( \mu v^\mu \right) =0\qquad \eta _\mu ^\nu
\nabla _\nu \,\left( nu^\mu \right) =0  \label{cur}
\end{equation}
where projective tensors are 
\begin{equation}
\perp _\sigma ^\mu =g_\sigma ^\mu -\eta _\sigma ^\mu \qquad \eta ^{\mu \nu
}=v^\mu v^\nu -u^\mu u^\nu  \label{fund}
\end{equation}
and $u^\mu $ and $v^\mu $ are unit vectors 
\begin{equation}
u^\mu u_\mu =-1=-v^\mu v_\mu \qquad u^\mu v_\mu =0  \label{vectors}
\end{equation}
Parameters $U$, $T$, $\mu $, $n$ are determined by the EOS, and \cite
{CP95,Carter2000} 
\begin{equation}
\mu ^2=\chi \qquad n^2=K^2\chi \qquad \chi >0  \label{x1}
\end{equation}
at spacelike currents, while 
\begin{equation}
\mu ^2=-K^2\chi \qquad n^2=-\chi \qquad \chi >0  \label{x2}
\end{equation}
at timelike currents, where function $K\left( \chi \right) $ is ultimately
defined as \ 
\begin{equation}
K=-\left( 2\frac{d\Lambda }{d\chi }\right) ^{-1}  \label{k}
\end{equation}

The extrinsic equations of motion (\ref{ex1}) admit solutions in the form of
infinitesimal perturbations of the worldsheet (''wiggles''), which propagate
at velocity \cite{Carter89a}\textrm{\ } 
\begin{equation}
c^{\perp 2}=\frac TU=\left( \frac{\Lambda +\chi /K}\Lambda \right) ^{\mathrm{%
sign}\chi }  \label{sounde}
\end{equation}
The 'intrinsic' equations of motion (\ref{cur}) admit infinitesimal
longitudinal perturbations (''woggles'' or sound waves) which propagate
within the worldsheet at velocity\textrm{\ } 
\begin{equation}
c^2=-\frac{dT}{dU}=\frac n\mu \frac{d\mu }{dn}=\left( 1+2\frac{K^{\prime
}\chi }K\right) ^{-\mathrm{sign}\chi }  \label{soundl}
\end{equation}
where\ 
\begin{equation}
K^{\prime }=\frac{dK}{d\chi }  \label{kk}
\end{equation}

The Lagrangian $\Lambda $ is obtained by numerical integration over the
coordinates orthogonal to the worldsheet. However, there are derived a few
explicit analytical models. The linear model \cite{SPG87,VV87,CHT87} 
\begin{equation}
\Lambda =-m^2-\frac \chi 2  \label{lagr1}
\end{equation}
is applied to the cosmic strings, carrying fermionic currents. The following
models are applied to the cosmic strings which carry bosonic currents: 
\begin{eqnarray}
\Lambda &=&-m\sqrt{m^2+\chi }\qquad \mathrm{Ref.\,}\cite{NO87}  \label{lagr2}
\\
\Lambda &=&-m^2-\cfrac \chi 2\left( 1-\cfrac \chi {m_{*}^2}\right) \qquad 
\mathrm{Ref.\,}\cite{Carter2000}  \label{lagr3} \\
\Lambda &=&-m^2-\cfrac \chi 2\left( 1+\cfrac \chi {m_{*}^2}\right)
^{-1}\qquad \mathrm{Ref.\,}\cite{CP95}  \label{lagr4} \\
\Lambda &=&-m^2-m_{*}^2\ln \left( 1+\cfrac \chi {m_{*}^2}\right) \qquad 
\mathrm{Ref.\,}\cite{CP95,HC08}  \label{lagr5}
\end{eqnarray}

The linear EOS (\ref{lagr1}), is not enough to discover discontinuities
(shock waves) because the sound speed is constant and equal to the speed of
light 
\begin{equation}
c=1  \label{const}
\end{equation}
However, shock waves are observed when more complicated bosonic EOS (\ref
{lagr2}) is taken.

\begin{figure}[tbp]
\caption{The speed of extrinsic perturbations $c^{\perp }$ vs current $\chi $
(parameters $m=1$, $m_{*}=0.5$)}
\label{fig1}{\includegraphics[scale=0.6]{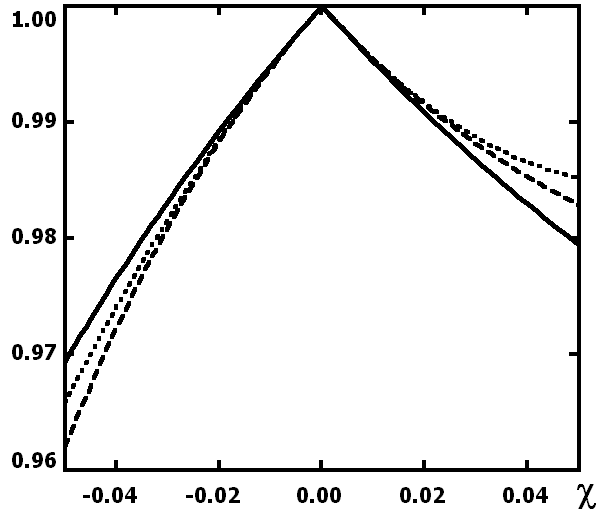}}
\par
Solid line -- model (\ref{lagr5}), dashed line -- model (\ref{lagr4}),
dotted line -- model (\ref{lagr3})
\end{figure}

A typical behavior of $c^{\perp }$ vs $\chi $ is shown in Fig. \ref{fig1}.
The speed $c^{\perp }(\chi )$ increases at $\chi <0$, and decreases at $\chi
>0$. In the ''transonic'' model (\ref{lagr2}) the speed of transversal
perturbations coincides with the sound speed 
\begin{equation}
c^{\perp }=c  \label{ctran}
\end{equation}
while for all other models (\ref{lagr3})-(\ref{lagr5}) it is always \cite
{CP95,Carter2000,HC08}: 
\begin{equation}
c^{\perp }>c  \label{ctran2}
\end{equation}

\section{Shock waves}

The sound perturbations (\ref{soundl}) are involved in the ''intrinsic''
equations of motion (\ref{cur}) and these perturbations can transform into
discontinuities similar to relativistic shock waves \cite{V99,MP00,CCMP02}.
The shock-wave solution must satisfy the stability criterion or evolutionary
condition \cite{CF59,Thorne73}, that for strings is formulated as \cite{V99}%
: 
\begin{equation}
w_{-}>c_{-}\qquad w_{+}<c_{+}  \label{evol}
\end{equation}
where labels ''$-$'' and ''$+$'' correspond to the state before and behind
the shock, respectively. This stability criterion (\ref{evol}) results in
inequality \cite{TV2011a} 
\begin{equation}
w_{-}>c_{-}>c_{+}>w_{+}  \label{ws1}
\end{equation}
for all string models (\ref{lagr2})-(\ref{lagr5}) in the ''magnetic''
regime. In the ''electric'' regime another inequality takes place 
\begin{equation}
c_{+}>w_{+}>w_{-}>c_{-}  \label{ws2}
\end{equation}

The evolutionary condition (\ref{evol}) also results in the growth of the
current \cite{TV2011a} 
\begin{equation}
\chi _{+}>\chi _{-}  \label{elres}
\end{equation}
However, the magnitude of shock wave in the ''electric'' regime (at $\chi
_{-}<0$) cannot exceed $\left| \chi _{-}\right| $ , so that a transition to
the ''magnetic'' regime $\chi _{-}<0\rightarrow \chi _{+}>0$ is impossible.
As for the magnitude of ''magnetic'' shock wave (at $\chi _{-}>0$), it can
be arbitrary.

Our preliminary analysis \cite{V99} was based on intuitive statement: if
arbitrary ''electric'' shock wave is not admitted, then, no ''electric''
shock is possible. However, the inequality (\ref{elres}) does not disqualify
small-amplitude shock waves in the ''electric'' regime. Nevertheless, no
shock wave was discovered in numerical simulations in the ''electric''
regime \cite{MP00, CCMP02}. The regime of timelike currents $\chi <0$\ can
admit only smooth solution instead of shock-wave discontinuity. Hence, there
is another physical mechanism which is making ''electric'' shocks
impossible. The puzzle is hidden in the shock wave instability caused by
perturbations of the string worldsheet.

\section{Instability to perturbations of worldsheet}

What happens with the string geometry when the intrinsic equations of motion
(\ref{cur}) admit discontinuous solution $\chi _{+}\neq \chi _{-}$? It is
clear that the speed of transversal perturbations (\ref{sounde}) is subject
to change, hence, $c_{+}^{\perp }\neq c_{-}^{\perp }$. The growth of current
(\ref{elres}), in the view of Fig. \ref{fig1}, will always result in 
\begin{equation}
c_{-}^{\perp }>c_{+}^{\perp }  \label{wse1}
\end{equation}
in the ''magnetic'' regime ($\chi >0$) of all models (\ref{lagr2})-(\ref
{lagr5}). The growth of current (\ref{elres}) in the ''electric'' regime ($%
\chi <0$) results in 
\begin{equation}
c_{-}^{\perp }<c_{+}^{\perp }  \label{wse2}
\end{equation}
Of course, there is no transversal discontinuity in the sense of shock wave,
but rather the string geometry is changed \cite{MP00, CCMP02}. However,
perturbations of the worldsheet may lead to instability of the shock front
which is called as corrugation instability in the mechanics of continuous
media \cite{LL87}.

\begin{figure}[tbp]
\caption{Shock wave velocities and velocities of extrinsic perturbations in
the reference frame co-moving the shock wave front}
\label{fig2}{\includegraphics[scale=0.5]{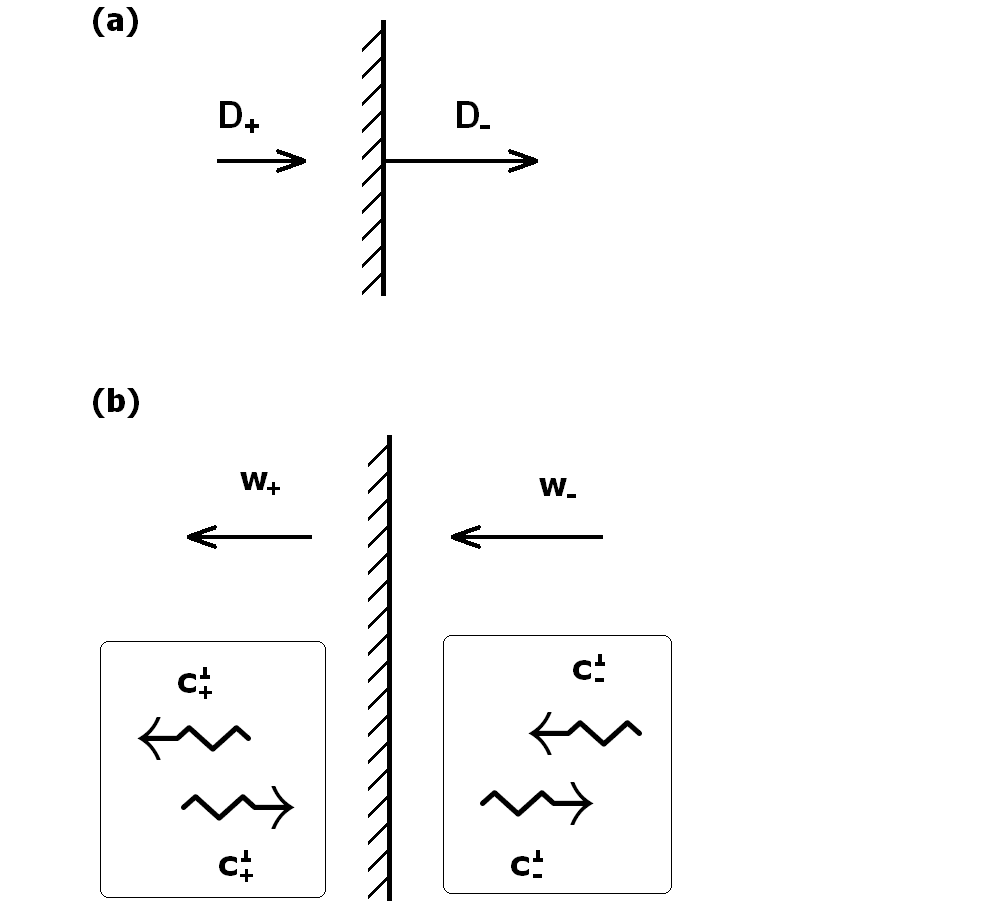}} %
%{(The state behind the shock is labeled by "$+$").}
\end{figure}

Consider a shock wave which propagates along the string at velocity $D_{-}$
(see Fig. \ref{fig2}a), and there is finite flow $D_{+}\neq 0$ behind the
shock front (while the shock is reduced to a sound wave in the limit $%
D_{+}\rightarrow 0$). Let us consider the problem in the reference frame,
co-moving the shock where the shock front is at rest, and the flow before
the front has velocity $w_{-}=-D_{-}$, while the flow behind the front has
velocity $w_{+}$ (positive direction is taken from the right to the left).
Extrinsic perturbations can appear before and behind the shock hypersurface
and they can propagate in two different directions with respect to the shock
front (see Fig. \ref{fig2}b).

\begin{figure}[tbp]
\caption{Possible relations between velocities of perturbations $w_-\oplus
c_- $ and $w_+\oplus c_+$ before and behind the front. Perturbations can be
emitted from the shock hypersurface (two bottom graphs). }
\label{fig22}{\includegraphics[scale=0.5]{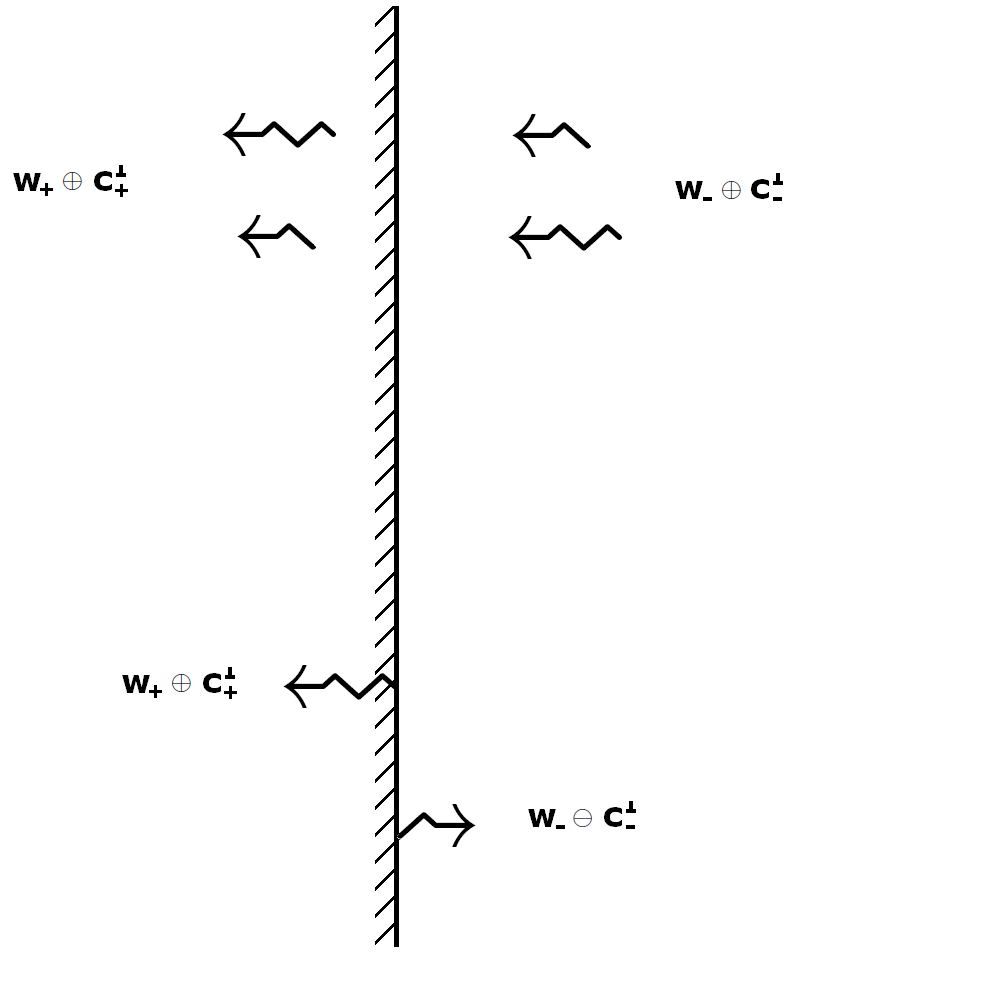}} %
%{(The state behind the shock is labeled by "$+$").}
\end{figure}

A perturbation, coming from infinity behind the front (see Fig. \ref{fig22}%
), runs at velocity 
\begin{equation}
C_{+}=w_{+}\oplus c_{+}^{\perp }=\frac{w_{+}+c_{+}^{\perp }}{%
1+w_{+}c_{+}^{\perp }}  \label{gof3}
\end{equation}
and before the front its velocity is 
\begin{equation}
C_{-}=w_{-}\oplus c_{-}^{\perp }=\frac{w_{-}+c_{-}^{\perp }}{%
1+w_{-}c_{-}^{\perp }}  \label{gof4}
\end{equation}
\textrm{\ }If $C_{-}^{\perp }<C_{+}^{\perp }$, then, the perturbation behind
the front becomes fully independent because no perturbation from the domain
before the shock will interfer with it. The perturbations before the shock
have no influence on the perturbations behind the shock, and the shock wave
hypersurface divides the string into two independent domains. To avoid it,
we must request\textrm{\ } 
\begin{equation}
\frac{w_{-}+c_{-}^{\perp }}{1+w_{-}c_{-}^{\perp }}>\frac{w_{+}+c_{+}^{\perp }%
}{1+c_{+}^{\perp }w_{+}}  \label{gof33}
\end{equation}

\begin{figure}[tbp]
\caption{Possible relations between velocities of perturbations $w_- \ominus
c_- $ and $w_+\ominus c_+$ before and behind the front. Perturbations
emitted from the shock hypersurface are depicted in the bottom). }
\label{fig222}{\includegraphics[scale=0.5]{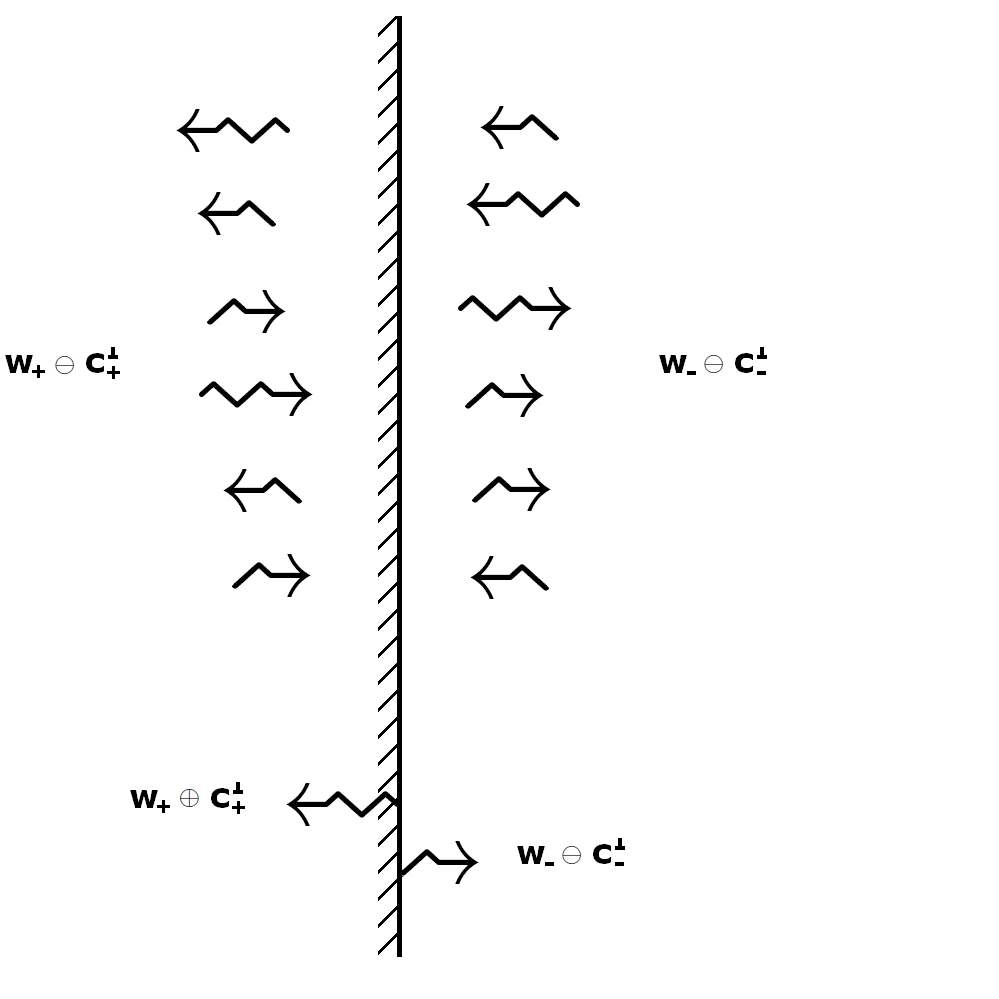}} %
%{(The state behind the shock is labeled by "$+$").}
\end{figure}

A perturbation, coming from infinity before the front (see Fig. \ref{fig222}%
), runs at velocity 
\begin{equation}
\bar C_{-}=w_{-}\ominus c_{-}^{\perp }=\frac{w_{-}-c_{-}^{\perp }}{%
1-w_{-}c_{-}^{\perp }}  \label{gof1}
\end{equation}
and a perturbation behind the shock runs at velocity 
\begin{equation}
\bar C_{+}=w_{+}\ominus c_{+}^{\perp }=\frac{w_{+}-c_{+}^{\perp }}{%
1-w_{+}c_{+}^{\perp }}  \label{gof2}
\end{equation}
All possible relations between $\bar C_{-}$ and $\bar C_{+}$ are plotted in
Fig. \ref{fig222}. Without regard of the particular sign of velocities, the
inequality $\bar C_{-}<\bar C_{+}$ always implies that there will be no link
between the perturbations before and behind the shock. To avoid this
situation, we must request 
\begin{equation}
\bar C_{-}>\bar C_{+}\qquad \Leftrightarrow \qquad \frac{w_{-}-c_{-}^{\perp }%
}{1-w_{-}c_{-}^{\perp }}>\frac{w_{+}-c_{+}^{\perp }}{1-w_{+}c_{+}^{\perp }}
\label{gof11}
\end{equation}
It may occur $\bar C_{-}<0$ and $\bar C_{+}<0$ (both perturbations propagate
from the left to the right and collinear to the shock velocity $D_{-}$),
however, as soon as $\bar C_{-}>\bar C_{+}$ or $\left| \bar C_{-}\right|
<\left| \bar C_{+}\right| $, then, a link between perturbations at oposite
sides of the front is established.

An extrinsic perturbation can be emitted by the shock in the direction of
its propagation, and it runs at velocity (\ref{gof1}). When an extrinsic
perturbation is emitted by the shock in the direction opposite to its
propagation, it runs at velocity (\ref{gof3}). Such perturbations are also
shown in Fig. \ref{fig22} and \ref{fig222}. If they propagate independent
from the shock flow, the stable shock structure will will be corrupted.

When either of inequalities (\ref{gof11}) or (\ref{gof33}) is borken, the
energy will be scattered beyond the self-consistent shock-wave regime that
implies instability and decay of the shock \cite{LL87}. The constraints (\ref
{gof11}) and (\ref{gof33}) are enough for existence of stable shock waves.

\begin{figure}[tbp]
\caption{Velocities $\bar C_{-}$ (\ref{gof1}) [solid] and $\bar C_{+}$ (\ref
{gof2}) [dashed] for EOS (\ref{lagr3}) at $m=m_{*}=1$ and various initial
current $\chi_{-}$ and increment $\Delta \chi$}
\label{fig3}{\includegraphics[scale=0.6]{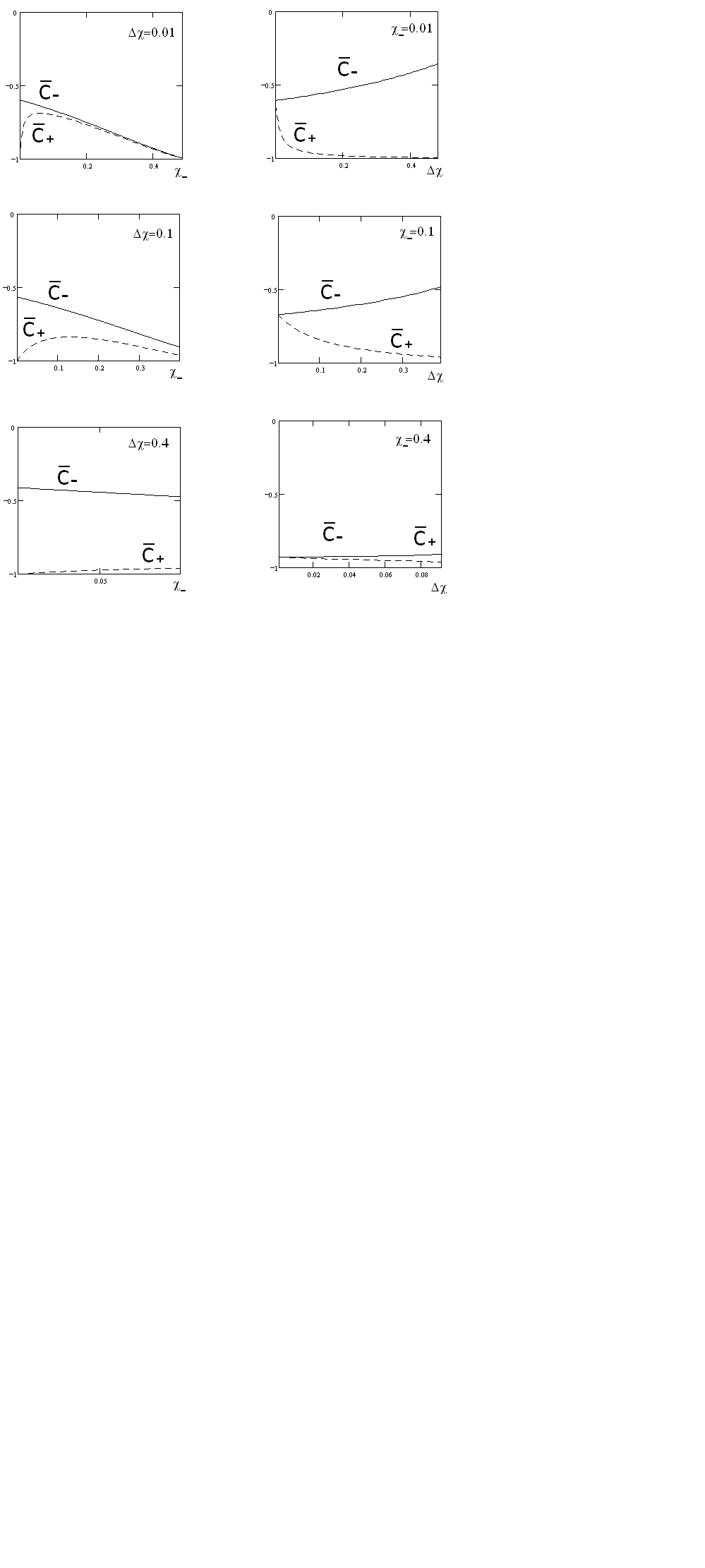}}
\end{figure}

\begin{figure}[tbp]
\caption{The same plots as in Fig. \ref{fig3} but for EOS (\ref{lagr4}) }
\label{fig33}{\includegraphics[scale=0.6]{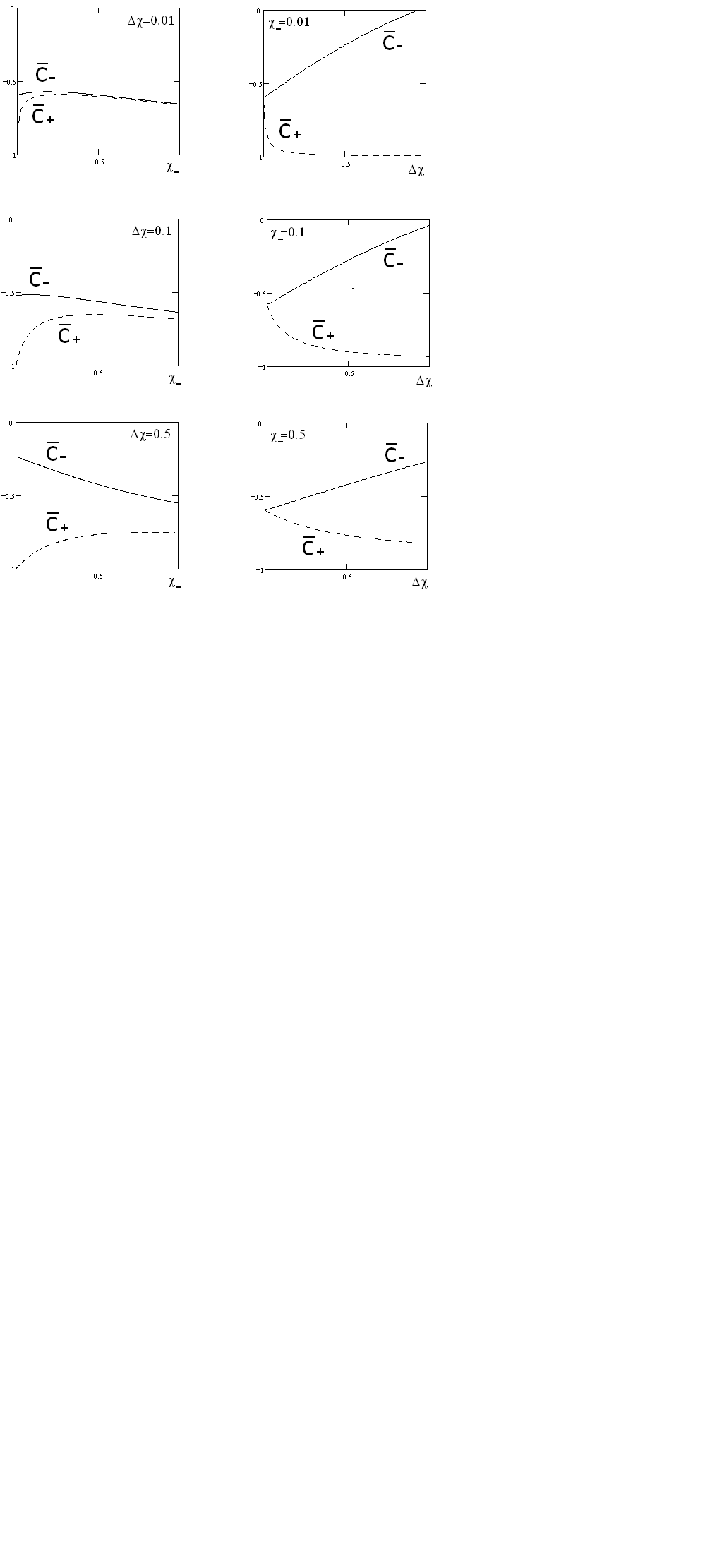}}
\end{figure}

\begin{figure}[tbp]
\caption{The same plots as in Fig. \ref{fig3} but for EOS (\ref{lagr5}) }
\label{fig333}{\includegraphics[scale=0.6]{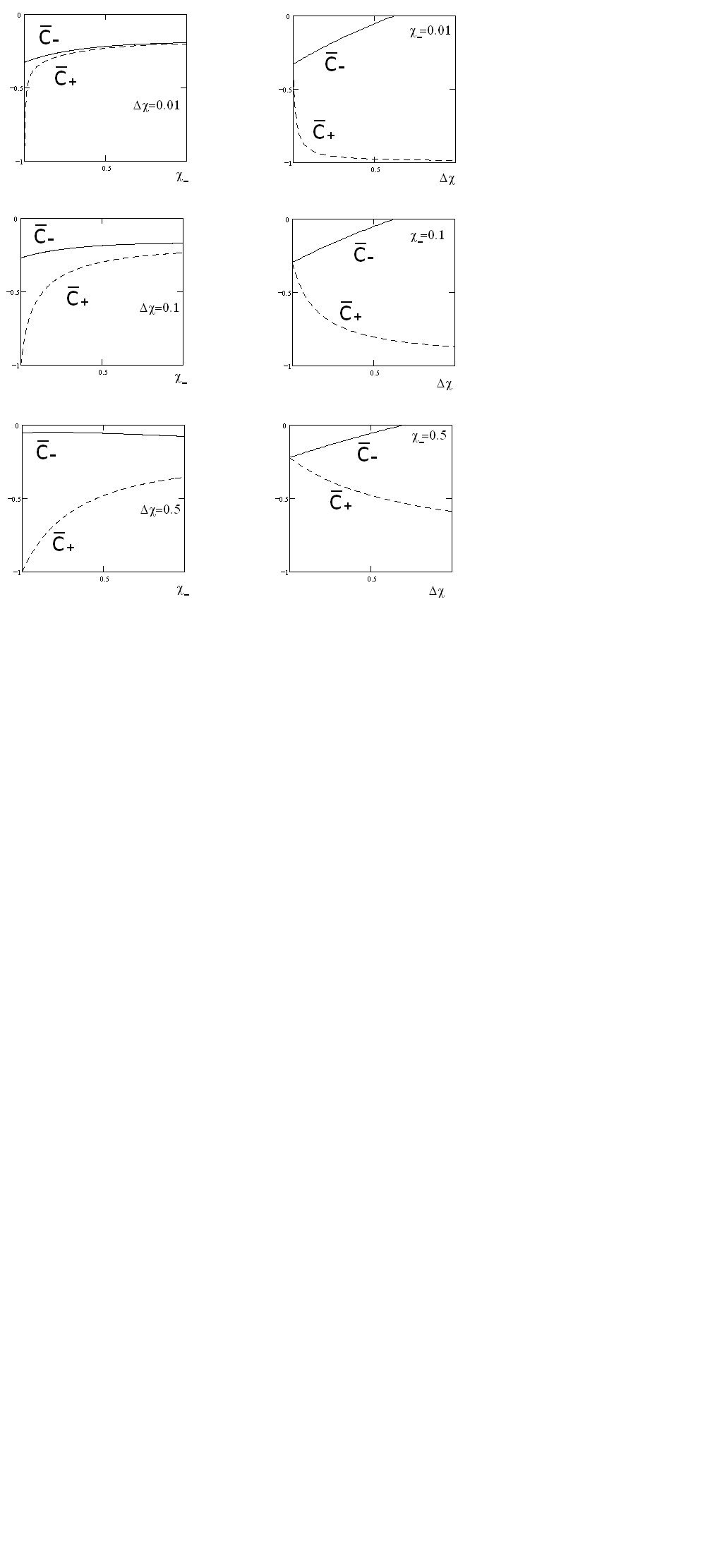}}
\end{figure}

In the light of (\ref{ws1}) and (\ref{wse1}), inequality (\ref{gof33}) is
always satisfied in the ''magnetic'' regime. Inequality (\ref{gof11}) is
automatically satisfied in the ''magnetic'' regime of the ''transonic''
model (\ref{lagr2}), as it follows from (\ref{ctran}) and (\ref{ws1}). As
for other three models (\ref{lagr3})-(\ref{lagr5}), the inequality (\ref
{gof11}) is also satisfied but it can be clarified in the direct calculation
(see Fig.~\ref{fig3}, \ref{fig33}, \ref{fig333}).

In the light of (\ref{ws2}) and (\ref{wse2}), the inequality (\ref{gof33})
is never satisfied in the ''electric'' regime, and it is the physical reason
why no ''electric'' shock was found in the numerical simulations \cite{MP00,
CCMP02}.

\section{Conclusion}

A superconducting cosmic string may admit a stable shock-like discontinuity
of the current when the latter is spacelike ($\chi >0$). A discontinuity of
timelike current ($\chi <0$) cannot not exist. In the present paper we have
explained why the ''electric'' shocks are impossible: it is due to the shock
instability to extrinsic perturbations of the string worldsheet. As soon as
this instability takes place, i.e. when either of inequality (\ref{gof33})
or (\ref{gof11}) is broken, the energy of the shock wave will dissipate. If
an arbitrary discontinuity $\Delta \chi \neq 0$ is created in the
''electric'' regime, e.g. during intercommutation of distinct string loops,
this discontinuity will be unstable, its energy will be converted without
restriction into extrinsic vibrations of the string worldsheet, and the
discontinuity will decay until it becomes a smooth transition $\chi
_{-}\Rightarrow \chi _{+}$.

As for the stable shock waves in the ''magnetic'' regime they do exist \cite
{V99,MP00,CCMP02}, and no restriction of their existence was found. However,
a discontinuity with initial state $\chi _{-}=0$ will be unstable because
the formula (\ref{soundl}) yields $c_{-}\left( \chi =0\right) =1$, and the
shock wave velocity $w_{-}=1$ does not satisfy the evolutionary condition (%
\ref{evol}). We can expect some minimal amplitude $\chi _{\min }>0$ to
trigger a discontinuity, and special analysis of ''magnetic'' shock wave at $%
\chi _{-}\rightarrow 0+$ is desirable. It is the subject for further
research.

%}

\end{document}